\documentclass{aa}  

\usepackage{graphicx}
\usepackage{tabularx}
\usepackage{color}

\usepackage{txfonts}
\newcommand{\ysz}{Y$_{\textrm{SZ}}$}

\newcommand{\planck}{{\it Planck}}
\newcommand{\chexmate}{{CHEX-MATE}}
\newcommand{\euclid}{{\it Euclid}}
\newcommand{\chandra}{{\it Chandra}}
\newcommand{\rosat}{{\it ROSAT}}
\newcommand{\xmm}{{XMM-{\it Newton}}}
\newcommand{\yclus}{\ensuremath{Y_{5R500}}}
\newcommand{\rclus}{\ensuremath{R_{500}}}
\newcommand{\thetaclus}{\ensuremath{\theta_{500}}}
\newcommand{\mmmf}{M$_{\textrm{MMF3}}$}
\newcommand{\degree}{$^{\circ}$}
\newcommand{\mwl}{M$_{\text{WL}}$}
\newcommand{\obias}{$1-b_{\chi}$} 
\newcommand{\wlbias}{$1-b_{\text{WL}}$}

\usepackage{natbib}
\usepackage{subcaption}
\usepackage{tabularx}
\begin{document}

   \title{CHEX-MATE: The Impact of Triaxiality and Orientation on Planck SZ
Cluster Selection and Weak Lensing Mass Measurements}

   \author{H. Saxena \inst{1} \and
            J. Sayers \inst{1} \and
            A. Gavidia \inst{1} \and
            J.-B. Melin \inst{2} \and
            E.~T. Lau \inst{3} \and
            J. Kim \inst{4} \and
            L. Chappuis \inst{2, 17} \and 
            D. Eckert \inst{17} \and 
            S. Ettori \inst{6,7}  \and 
            M. Gaspari \inst{12} \and 
            F. Gastaldello \inst{8} \and
            S. Kay \inst{17} \and 
            L. Lovisari \inst{8,9} \and
            F. Oppizzi \inst{11} \and 
            M.D. Petris \inst{13} \and 
            G.W. Pratt \inst{5} \and
            E. Pointecouteau \inst{10} \and 
            E. Rasia \inst{14,15,16} \and
            M. Rossetti \inst{8} \and
            M. Sereno \inst{6,7}
}

   \institute{California Institute of Technology, 1200 East California Boulevard, Pasadena, California, USA
        \and 
            IRFU, CEA, Université Paris-Saclay, 91191 Gif-sur-Yvette, France
            \and 
            Smithsonian Astrophysical Observatory, Cambridge, Massachusetts, USA \and
            Department of Physics, Korea Advanced Institute of Science and Technology (KAIST), 291 Daehak-ro, Yuseong-gu, Daejeon 34141, Republic of Korea \and
            Université Paris-Saclay, Université Paris Cité, CEA, CNRS, AIM, 91191 Gif-sur-Yvette, France \and 
            INAF - Osservatorio di Astrofisica e Scienza dello Spazio di Bologna, via Piero Gobetti 93/3, I-40129 Bologna, Italy \label{inaf-oas} \and 
            INFN, Sezione di Bologna, viale Berti Pichat 6/2, I-40127 Bologna, Italy \label{infn-bo} \and
            INAF, Istituto di Astrofisica Spaziale e Fisica Cosmica di Milano, via A. Corti 12, 20133 Milano, Italy \and
            Center for Astrophysics $|$ Harvard $\&$ Smithsonian, 60 Garden Street, Cambridge, MA 02138, USA \and
            IRAP, CNRS, Université de Toulouse, CNES, Toulouse, France \and
            INFN-Sezione di Genova, Via Dodecaneso 33, 16146, Genova, Italy \and 
            Department of Physics, Informatics and Mathematics, University of Modena and Reggio Emilia, 41125 Modena, Italy \and 
            Dipartimento di Fisica, Sapienza Università di Roma, Piazzale Aldo Moro 5, I-00185 Rome, Italy \and
            INAF, Osservatorio Astronomico di Trieste, via Tiepolo 11, I-34131, Trieste, Italy \and 
            IFPU, Institute for Fundamental Physics of the Universe, Via Beirut 2, 34014 Trieste, Italy \and 
            Department of Physics; University of Michigan, Ann Arbor, MI 48109, USA \and 
            Department of Astronomy, University of Geneva, Ch. d’Ecogia 16, CH-1290 Versoix, Switzerland \and 
            Jodrell Bank Centre for Astrophysics, Department of Physics and Astronomy, The University of Manchester, Manchester M13 9PL, UK}
   \date{}

 
  \abstract
   {Galaxy cluster abundance measurements are a valuable tool for constraining cosmological parameters like the mass density ($\Omega_m$) and density fluctuation amplitude ($\sigma_8$). Wide area surveys detect clusters based on observables,  such as the total integrated Sunyaev-Zel'dovich effect signal (\ysz) in the case of \planck.
   Quantifying the survey selection function is necessary for a cosmological analysis, with completeness representing the probability of detecting a cluster as a function of its intrinsic properties such as \ysz\ and angular scale $\theta_{500}$.} 
   {The completeness of the \planck-selected \chexmate\ cluster catalog is determined from mock observations of clusters with triaxial shapes and random orientations with physically-motivated distributions of axial ratios. From these mocks, the distribution of shapes and orientations of the detected clusters, along with any associated bias in weak lensing-derived mass (\mwl) due to this orientation-dependent selection, denoted as \obias, is obtained.}  
   {Employing a Monte-Carlo method, we inject triaxial cluster profiles into random positions within the \planck\ all-sky maps, and subsequently determine the completeness as a function of both geometry and SZ brightness. This is then used to generate 1000 mock \chexmate\ cluster catalogs. \mwl\ is computed for these mock \chexmate\ clusters, and for equal-sized samples of randomly selected clusters with similar mass and redshift distributions.}
   {Cluster orientation impacts completeness, with a higher probability of detecting clusters elongated along the line of sight (LOS). This leads to \obias\ values of 0--4\% for \chexmate\ clusters relative to a random population. The largest increase in \mwl\ is observed in the lowest mass objects, which are most impacted by orientation-related selection bias.} 
   {
   Clusters in \planck\ SZ-Selected catalogs are preferentially elongated along the LOS and have an on-average bias in \mwl\ relative to randomly selected cluster samples. This bias is relevant for upcoming SZ surveys like CMB-S4, and should be considered for surveys utilizing other probes for cluster detection, such as \euclid.}


\titlerunning{Cluster Triaxiality and SZ selection}
\authorrunning{H. Saxena et al. }
   \maketitle
%

\section{Introduction} \label{introduction}
Galaxy cluster abundance is a powerful cosmological probe, offering incisive tests of the $\Lambda$CDM model and precise measurements of fundamental cosmological parameters \citep[e.g.,][]{Voit2005,Allen2011}. As the most massive gravitationally bound structures in the universe, galaxy clusters trace the late stages of structure formation, representing the evolution of the highest peaks in the primordial matter density field. Their abundance is particularly useful for constraining both the mean matter density ($\Omega_m$) and the amplitude of matter fluctuations ($\sigma_8$), defined as the root mean square of linear density perturbations on scales of $8~\text{h}^{-1}$ Mpc. Recent studies using cluster catalogs selected from X-ray, optical, and millimeter wavelength surveys have yielded increasingly precise constraints on these parameters \citep{Ghirardini_2024,DESY3_2025_Halo_Abundance,Bocquet2024}, although challenges related to halo mass calibration and selection biases still remain.
\\\\
The hot intracluster medium (ICM) emits X-rays via thermal Bremsstrahlung \citep{Sarazin1986}, producing a bright diffuse signal that makes X-ray surveys a valuable tool for detecting galaxy clusters. Early X-ray halo abundance studies were based on catalogs of approximately 100 clusters detected in the \rosat\ All-Sky Survey (RASS), utilizing deep X-ray followup from \chandra\ to determine halo masses via hydrostatic equilibrium-based estimates \citep{2009Chandra_cosmo, Mantz2010}. Because the bias associated with such mass estimates is difficult to quantify, later work improved on these results by utilizing weak lensing (WL) observations of a sub-sample of clusters to calibrate the halo mass scale \citep{Mantz_2015}. The more recent XXL survey from \xmm, which is much narrower and deeper than the RASS, delivered a comparably-sized cluster sample with WL again calibrating the mass scale for a halo abundance measurement \citep{Pacaud_2018}. The current state-of-the-art in X-ray cluster cosmology was obtained from the eROSITA All-Sky Survey (eRASS), based on over 5,000 X-ray-selected clusters \citep{Ghirardini_2024}. Cluster masses were calibrated using WL, in this case utilizing WL mass (\mwl) estimates from three different optical surveys \citep{Chiu2022,Kleinebreil2025,Grandis2024}, enabling a robust scaling between X-ray observables and total halo mass \citep[see also][]{Okabe2025}.
\\\\
In the optical regime, cluster detection is most often based on searching for overdensities of galaxies, for example via algorithms that attempt to identify a red sequence of cluster-member galaxies \citep[e.g.,][]{Rykoff2014}. For instance, commensurate with the RASS-based studies noted above the maxBCG cluster sample from the Sloan Digital Sky Survey (SDSS) was used by \citet{Rozo_Opt_cosmo} to derive cosmological constraints from over $10^4$ clusters and groups at low redshifts. 
To connect the detection observable, optical richness, with the underlying halo mass, stacked \mwl\ values within bins of fixed richness were used. Subsequent work by \citet{SDSS_cosmo_opt} combined abundance and WL measurements from the SDSS DR8 to jointly constrain cosmology and the richness–mass relation for a photometrically selected sample of redMaPPer clusters. Illustrating the challenges related to selection and mass calibration, the Dark Energy Survey (DES) released its initial halo abundance measurement soon after \citep{DES_2020_halo_abundance}, with an anomalously low value of the matter density that was later found to be due to biases in \mwl\ values for the lowest richness clusters \citep{DES_2021_halo_abundance,DES_SPT_2021}. The current state-of-the-art in optical cluster cosmology was presented in the DES Y3 results and the Third Release of the  Kilo-Degree Survey \citep{DESY3_2025_Halo_Abundance, KIDS_optical}, who find cosmological parameter values that are fully consistent with those obtained from primary CMB anisotropy measurements from \planck. 
\\\\
Scattering of CMB photons with electrons in the ICM produces an on-average boost in photon energy known as the Sunyaev-Zel'dovich (SZ) effect \citep{Sunyaev}, which allows for the detection of clusters in millimeter-wave surveys. 
Recent large area surveys from the Atacama Cosmology Telescope \citep{Hilton2021}, the South Pole Telescope \citep{Bleem2015,Bleem2024,Kornoelje2025}, and \planck\ \citep[hereafter P16]{PSZ2} have now provided catalogs with 1000s of clusters which have in turn been used to constrain cosmological parameters \citep{Sehgal2011,Bocquet2024,PSZ2Cosmo}.
Notably, the SZ effect surface brightness is independent of redshift, making these surveys approximately mass-limited across most redshifts and thus providing catalogs that are particularly attractive for halo abundance measurements. 
However, results from these surveys have generally been limited by systematic uncertainties on halo mass calibration. For example, \planck\ detected fewer clusters than predicted by the base $\Lambda$CDM model constrained by its primary CMB anisotropy measurement when hydrostatic mass estimates are used for the cluster measurement \citep{PSZ2Cosmo, Planck2018}. This motivated numerous WL studies to improve the mass calibration \citep[e.g.,][]{vonderLinden2014,Hoekstra2015}, although most of these calibrations still suggest at least mild tension between the cluster and CMB measurements.
\\\\
As noted above, a central challenge in cluster cosmology, both for halo abundance and clustering, is accurately calibrating the scaling relations between mass and observables---such as X-ray brightness, optical richness, and SZ flux (\ysz)---which are needed to connect cluster detections based on these observables to the Halo Mass Function (HMF). In general, systematic uncertainties in this calibration limit the precision of the derived cosmological constraints \citep[e.g.,][]{Pratt2019}. 
To improve mass calibration, nearly all recent efforts have focused on \mwl, which is sensitive to the total projected mass distribution and independent of the dynamical state of the ICM.
However, \mwl\ values have significant cluster-to-cluster scatter, along with an on-average bias, primarily from projection effects due to halo asphericity, orientation, and miscentering \citep[e.g.,][]{Meneghetti}. To quantify this scatter and bias, it is common to perform mock WL analyses of simulated clusters with random orientations \citep[e.g.,][]{euclid}. A limitation to such studies arises if the observable used to detect the cluster sample is sensitive to halo orientation, resulting in a selected population with a preferred geometry, in which case the scatter and bias in \mwl\ could be mis-estimated. 
\\\\
Understanding the selected population of galaxy clusters requires quantifying survey completeness, which measures the probability of detecting a cluster based on its observables. For the \planck\ SZ survey, completeness, $\chi(\yclus,\, \thetaclus,\, l, b)$, is defined as a function of four cluster parameters, where \yclus\ is the integrated SZ signal within 5\rclus, \thetaclus\ is the corresponding angular scale, and $(l, b)$ are the cluster’s galactic coordinates (P16)\footnote{\rclus\ denotes the spherical radius enclosing an average overdensity 500 times larger than the critical density of the universe at the cluster redshift}. Detectability is assessed using the observed signal-to-noise ratio (S/N) obtained from, e.g., the Multi-Matched Filtering (MMF3) algorithm \cite[P16, see also][]{Melin_2006, MMF3_og_paper}, which must exceed a threshold value. A robust Monte Carlo (MC) method for characterizing completeness involves injecting a large number of model clusters with a range of values for \yclus\ and \thetaclus\ into random locations within the \planck\ maps, and then computing the S/N from the MMF3 for each injected cluster. The \planck\ team did not consider asphericity in this calculation, and only injected clusters based on spherically symmetric models. Subsequent work by \citet{Gallo2024} estimated the \planck\ completeness using simulated clusters with complicated aspherical shapes more closely resembling reality, finding that steeper SZ signal profiles result in a higher probability of detection. The impact of halo ellipticity was found to be minimal. Overall, the change in completeness obtained by \citet{Gallo2024} results in approximately $1\sigma$ shifts in the derived values of $\Omega_m$ and $\sigma_8$ relative to the nominal \planck\ result.

In this work, we provide a formalism to investigate how halo triaxiality and orientation impacts cluster detection and SZ selection, focusing on the \planck-selected \chexmate\ sample. By utilizing smooth cluster models, rather than the simulated clusters of \citet{Gallo2024}, we are able to isolate the impact of triaxiality and orientation from other characteristics such as sub-structure and dynamical state. In addition, because our study is not limited to a relatively small number of simulated objects, we are also able to obtain sufficient statistics to characterize sub per cent level biases. The Cluster HEritage project with \xmm\ – Mass Assembly and Thermodynamics at the Endpoint of structure formation \citep{CHEXMATE} is an \xmm\ Heritage Project designed to study ICM physics and improve cluster-based cosmological constraints by refining links between observables and halo mass\footnote{\url{https://xmm-heritage.oas.inaf.it}}. It includes 118 clusters, divided into a volume-limited sample (Tier 1, $0.05<z<0.2$) and a mass-limited sample (Tier 2, $z<0.6$, $M>7.25 \times 10^{14} M_\odot$), each containing 61 clusters with four clusters in common. Further, all clusters in the catalog have a \planck\ $\text{S/N} > 6.5$, and Tier 1 is restricted to the northern sky, $\text{decl.} > 0$\degree. This deep, uniform, high-quality multi-probe dataset enables precise measurements of triaxial 3D cluster shapes and orientations \citep{Kim_2024}. Our goal is to model the SZ selection function of the \chexmate\ sample while incorporating cluster triaxiality through informative priors. The end result is a \chexmate\ completeness estimate that depends on triaxial geometry, along with a determination of the expected distribution of geometries for the \chexmate\ sample. We further explore the impact of this selection on the values of \mwl\ derived for the \chexmate\ clusters.
\\\\
This paper is organized as follows. In Section \ref{MMF3}, we describe the details of the MMF3 algorithm used to compute the detection and selection of clusters. In Section \ref{Triax}, we describe the triaxial model assumed for the ICM and total mass distribution of the clusters. In Section \ref{selec_func}, we describe the MC injection and detection to calculate the completeness for a set of both spherical and triaxial clusters, and quantify how the completeness depends on the triaxial shape and orientation parameters. In Section \ref{mock_CHEX}, we discuss the algorithm for generating a mock \chexmate\ catalog, starting from assembling a mock cluster sample from a HMF, injecting triaxial clusters from this sample based on informative priors on their shape and orientation, and subsequently detecting these mock clusters using the MMF3 algorithm. In Section \ref{WL_mass_bias}, we estimate \mwl\ of these mock \chexmate\ samples to quantify any WL mass bias, defined as \wlbias\ $= M_{\text{WL}}/M_{\text{true}}$ due to preferentially selected shapes and orientations. Finally, in Section \ref{conc}, we summarize the main results of this work. 

\section{Multi-Matched Filtering Algorithm (MMF3)} \label{MMF3}
The \chexmate\ sample was selected from the \planck\ PSZ2 catalog, and we thus follow P16 in their construction of the MMF3. We have independently recreated the MMF3 algorithm, with an implementation as close as possible to that used by \planck, with our verification of this implementation detailed below. 

The total signal in the \planck\ maps at each of the High Frequency Instrument (HFI) frequencies at a given position in the sky is modeled as 
\begin{equation}
    \boldsymbol{m(x)} = y_0 \boldsymbol{t_{\theta_s}(x)} + \boldsymbol{n(x)},
\end{equation}
where $\boldsymbol{x}$ is the 2D pixel coordinate within the map, $\boldsymbol{n(x)}$ is the unwanted astrophysical signals and instrumental noise, and $\boldsymbol{t_{\theta_s}(x)}$ is the SZ signal from a cluster with angular size $\theta_s$ and Comptonization parameter $y_0 = k\sigma_T\int n_e T_e dl /(m_e c^2) $. To obtain this SZ signal profile, we construct the normalized cluster profile $\tau_{\theta_s}$, which is described by the Generalized Navarro-Frank-White (GNFW) pressure profile \citep{Nagai_2007,Arnaud_2010}
\begin{equation}
    \tau(X) = \frac{P_0}{(c_{500}*X)^\gamma [1+(c_{500}*X)^\alpha]^{(\beta-\gamma)/\alpha}},
    \label{eqn:gnfw}
\end{equation}
with $P_0$ giving the amplitude, $c_{500} = 1.177$, $\alpha = 1.0510$, $\beta = 5.490$, and $\gamma = 0.3081$. Here, $X=\theta/\theta_{500}$, where $\theta$ is the angular radius and $\theta_s = \theta_{500}/c_{500}$. This 3D profile is then integrated along the LOS to give the projected 2D signal. This 2D map is then convolved with the \planck\ beam in each HFI channel, $b_i$, which are assumed to have Gaussian profiles as given in P16, with the final observed SZ signal given by 
\begin{equation}
    \boldsymbol{t_{\theta_s}(x)}_i = j_{\nu}(\nu_i)[b_i*\tau_{\theta_s}](\boldsymbol{x}).
\end{equation}
The spectral shape of the SZ signal is 
\begin{equation}
    j_\nu(x_\nu) = x_\nu T_{\text{CMB}} \left[\frac{e^{x_\nu}+1}{e^{x_\nu}-1} - 4\right],
\end{equation}
where $x_\nu = \frac{h\nu}{k_B T_{\text{CMB}}}$, the dimensionless frequency. Following P16, we do not consider the effect of relativistic corrections to the SZ effect here. For MMF3, we apply the detection algorithm on square cutouts of the all-sky maps centered on the cluster position with sides of 2.5\degree. The noise matrix is estimated directly from these local maps, by mean subtracting the maps, applying a Hanning window to ensure periodicity at the map edges, and then computing the cross-power spectra for the $6 \times 6$ \planck\ HFI frequencies. An azimuthal averaging procedure is applied, and then this cross-channel matrix is inverted for each Fourier-space pixel $\boldsymbol{k}$, defining the inverse noise covariance matrix $\boldsymbol{P^{-1}(k)}$. The uncertainty on the SZ signal detection is then estimated from 
\begin{equation}
    \sigma_{\theta_s} = \left[ \sum_k \boldsymbol{t_{\theta_s}(k)} \boldsymbol{P^{-1} (k)} \boldsymbol{t_{\theta_s}(k)} \right]^{-1/2}.
\end{equation}
The Matched Multi-filter assumes the spatial and spectral form for the SZ signal defined above. The filter ($\boldsymbol{\Psi_{\theta_s}}$) is uniquely specified by demanding a minimum variance estimate \citep{Melin_2006}, and constructed as 
\begin{equation}
    \boldsymbol{\Psi_{\theta_s}(k)} = \sigma_{\theta_s}^2 \boldsymbol{P^{-1} (k)} \boldsymbol{t_{\theta_s}(k)}.
\end{equation}
These are circularly symmetric filters, and show typical Fourier ringing as demonstrated in \citet{Melin_2006} as a result of down-weighting the large angular scale modes contaminated with primary CMB anisotropy and dust, and use both spatial and frequency weighting to optimally extract the cluster signal from the maps. 
This allows us to recover an estimate of the signal, quantified by the central Comptonization parameter as 
\begin{equation}
    \hat{y}_0 = \sum_x \boldsymbol{\Psi_{\theta_s} (x)} \boldsymbol{m(x)},
\end{equation}
and thus finally measuring the S/N of the cluster as $\hat{y}_0/\sigma_{\theta_s}$. 

To verify our implementation of the MMF3 algorithm relative to the one utilized by P16, we compare our derived S/N values and integrated SZ signals to their published results for the \chexmate\ sample. To do this, we first obtain 2.5\degree~$\times$~2.5\degree\ cutouts centered on the MMF3 position reported in P16 for each \chexmate\ cluster from each of the six all-sky HFI channel \planck\ maps. $\Psi_{\theta_s}$ is then applied to each of these maps, with the angular scale of the filter spanning $\theta_s$ values between 0.8\arcmin\ to 32\arcmin. We define the cluster size as the filter scale that maximizes the S/N at the location of the cluster and the best-fit SZ signal is given by the corresponding $\hat{y}_0$ parameter. The integrated signal within a radial extent of 5\rclus\ is then defined as 
\begin{equation}
    Y_{5R500} = \hat{y}_0 \int_{r<5R_{500}} \tau_{\theta_s} (r) r^2 dr.
\end{equation}

\begin{figure}[!tbp]
  \centering
    \includegraphics[width=0.49\textwidth]{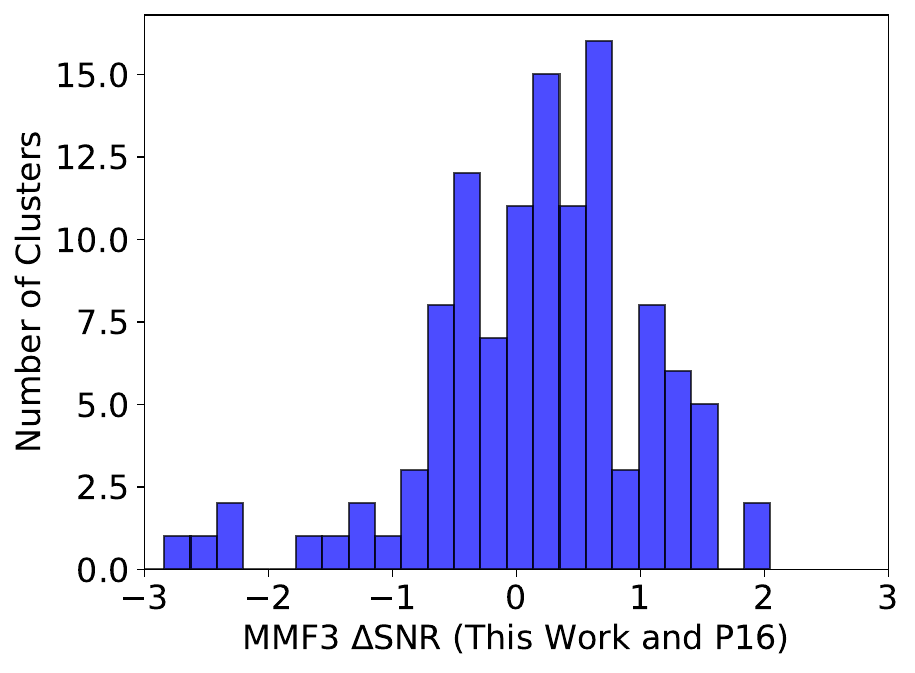}
  \caption{Histogram of the difference in S/N obtained from our implementation of the MMF3 detection algorithm and the one utilized by P16 for the \chexmate\ clusters.}
  \label{fig:S/N_y_comparison}
\end{figure}

Figure \ref{fig:S/N_y_comparison} shows a comparison of our derived values of S/N for the \chexmate\ sample to those obtained from P16. 
On average, we find very good agreement, with slight differences due to inherent instabilities in the MMF3 algorithm. At large angular scales, corresponding to small values of $| \boldsymbol{k} |$, the primary CMB anisotropy dominates the lower frequency HFI maps. Since this signal is coherent across frequencies, it introduces strong inter-channel correlations in the noise. Similarly, in the high frequency HFI maps, Galactic dust emission becomes the dominant signal and this emission is also spatially coherent and highly correlated across these frequencies. 
Due to these large correlations, even minor differences in how the noise is modeled can propagate nonlinearly when inverting the noise covariance matrix, resulting in measurable discrepancies in the final S/N. These discrepancies can be quantified by the difference between S/N values obtained from our MMF3 implementation and the one utilized by P16. For the \chexmate\ sample, the median difference is $0.21$, with a standard deviation of $0.72$, see Fig.~\ref{fig:S/N_y_comparison}. Comparing to a Gaussian distribution with this mean and scale factor, a KS test yields a $p$-value of $0.56$, indicating that the observed difference in S/N is consistent with such a distribution. We note that this level of agreement is approximately a factor of two better than that found between the detection algorithms described in P16, suggesting that our implementation of the MMF3 is consistent with the one utilized by P16. Furthermore, applying our MMF3 algorithm results in only 3 out of 118 \chexmate\ clusters falling below the detection threshold of $\text{S/N} > 6.5$. As a result, we expect the overall impact on sample selection from our MMF3 implementation relative to that of P16 to be minimal.

\section{Triaxial Cluster Shapes} \label{Triax}
\begin{figure}[!tbp]
  \centering
    \includegraphics[width=0.49\textwidth]{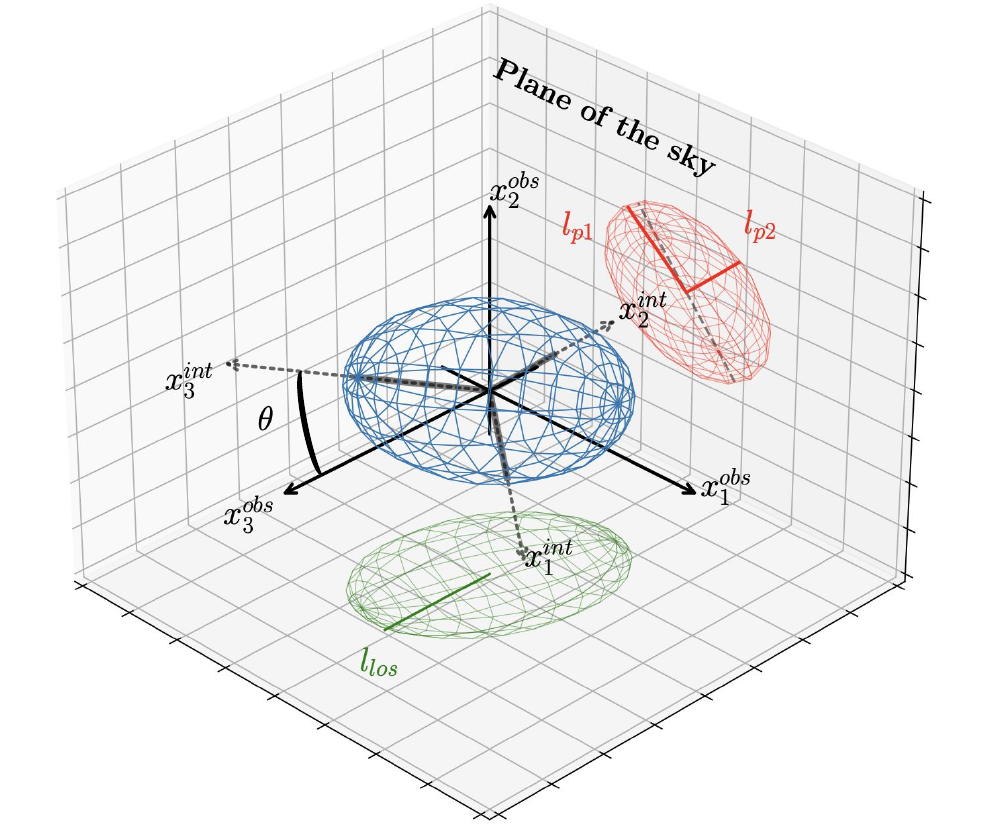}
  \caption{The triaxial ellipsoid model used in the analysis. The intrinsic coordinate system of the ellipsoid is shown by the dotted arrows, whereas the black arrows correspond to the observer’s coordinate system. The red ellipse shows the projection of the ellipsoid on the sky plane, with $l_{p1}$ as its semi-major axis. The green ellipse is the projection of the ellipsoid onto the plane that is perpendicular to the sky plane, and $l_{los}$ is the half size of the ellipse along the observer line of sight \citep{Kim_2024}. }
  \label{fig:triax_ellip}
\end{figure}

The matter distribution within galaxy clusters is better approximated by a triaxial ellipsoid than a sphere \citep{Despali2014,Bonamigo2015,Vega-Ferrero2017}, primarily driven by the anisotropic accretion of matter along cosmic filaments and the complex dynamics of structure formation. Simulations suggest that modeling clusters as 3D triaxial ellipsoids results in lower bias and less intrinsic scatter in estimating intrinsic quantities \citep{Becker_2011}.
Observations also strongly support the triaxial nature of galaxy clusters. Ellipticity has been detected in a variety of observational probes: X-ray surface brightness maps reveal elongated morphologies \citep{Lau_X_ray_2012}, SZ maps show asymmetric pressure distributions \citep{Sayers_SZ_ellip}, and weak lensing reconstructions uncover elliptical mass distributions \citep{Oguri_WL_ellip}, among others. These complementary data sets highlight the motivation for 3D triaxial modeling in both theoretical and observational contexts. 
As some examples of such modeling, \citet{Limousin_2013} present a general parametric framework within a triaxial basis to simultaneously fit complementary data sets from X-ray, SZ and lensing, and \citet{Clump3D} demonstrate cluster shape constraints from a multi-probe 3D analysis of the X-ray regular Cluster Lensing and Supernova survey with Hubble (CLASH) clusters from a combined analysis of strong and weak lensing, X-ray photometry and spectroscopy, and SZ.
\\\\
Following these works, we assume a geometry described by a triaxial ellipsoid with the ICM pressure following a distribution described by a radial profile in this triaxial basis, consistent with the formalism described in \citet{Kim_2024}. The main parameters involved in the construction of this profile are related to the intrinsic versus observed coordinate system of a triaxial ellipsoid. The projection of the ellipsoid on the sky plane is defined by an ellipse with a semi-major axis $l_{p1}$ and eccentricity $q_p$, while the projection of the ellipsoid along the line of sight is defined by a semi-axis $l_{los}$. $l_{p1}$, $q_p$, and $l_{los}$ have complicated descriptions in terms of the viewing angle and geometry of the ellipsoid $(q_1, q_2, \cos\theta, \phi, \psi)$, where $q_1, q_2$ are the minor-to-major and intermediate-to-major axial ratios of the ellipsoid, respectively, with the angles defined as the Euler angles between the intrinsic and observed coordinate system. These are shown in Fig. \ref{fig:triax_ellip} for clarity. 
\\\\
We model the pressure as a radial profile in the elliptical basis parameterized by the ellipsoidal radius $\zeta$, where $\zeta^2 = (x_1/q_1)^2 + (x_2/q_2)^2 + x_3 ^2$, which is used in place of the spherical radius in the GNFW model given in Eq. ~\ref{eqn:gnfw}. Within this framework, the 2D maps are calculated from the 3D ellipsoidal model projected onto the sky plane. The model Compton-y parameter is
\begin{equation}
y_{\text {model }}\left(x_{\xi} ; l_{\mathrm{p1}}\right)=\left(2 l_{\mathrm{p1}} e_{\|}\right)\left(\frac{\sigma_{\mathrm{T}}}{m_e c^2}\right) \int_{x_{\xi}}^{\infty} P_e\left(x_\zeta\right) \frac{x_\zeta}{\sqrt{x_\zeta^2-x_{\xi}^2}} d x_\zeta \text {, }
\end{equation}
where $y$ has been described in terms of the elliptical radius on the sky $x_\xi^2 = \left( x_1^2 + (x_2/q_p)^2 \right)\left(l_s/l_{p1}\right)^2 $, $x_\zeta = \zeta/l_s$, $q_p$ is the minor-to-major axial ratio of the observed projected isophote, $l_s$ is the semi-major axis of the ellipsoid, $e_{\|} = l_{los}/l_{p1}$ is the elongation parameter of the ellipsoid and $P_e$ is obtained from the GNFW profile
\begin{equation}
    P_e\left(x_\zeta\right)=\frac{P_0 P_{500}}{\left(c_{500} x_\zeta \frac{l_s}{R_{500}}\right)^{\gamma_p}\left[1+\left(c_{500} x_\zeta \frac{l_s}{R_{500}}\right)^{\alpha_p}\right]^{\left(\beta_p-\gamma_p\right) / \alpha_p}}
    \label{eq:p_e_zeta}
\end{equation}
with the same GNFW profile parameters as used in Eq. \ref{eqn:gnfw}. Thus, a 3D triaxial model of a cluster is generated and then projected into a 2D map based on its geometrical shape parameters $ (q_{1}, q_{2}, \cos\theta, \phi, \psi) $ and its pressure given by the GNFW model.

We define the total SZ signal \yclus\ and angular extent \thetaclus\ for a triaxial cluster by computing the spherically-averaged radial profile of the 3D triaxial model. The spherical GNFW model is then fit to this spherically-averaged profile model to determine the best-fit spherical \thetaclus, which then defines the angular scale of that cluster. Similarly, \yclus\ ($Y_{500}$) is computed by integrating this spherically-averaged pressure profile within 5\thetaclus \ (\thetaclus). With this procedure, we ensure that the total integrated SZ signal \yclus\ is the same regardless of triaxial geometry. Because we define the elliptical radius relative to the major axis, the spherically-averaged \thetaclus\ is always smaller than its elliptical equivalent in Eq.~\ref{eq:p_e_zeta}.

\section{Selection Function} \label{selec_func}
Estimating the \planck\ completeness requires introducing realistic noise into mock observations of cluster models with known properties, such as \yclus\ and \thetaclus. To obtain representative noise, we directly sample the \planck\ all-sky maps, and then apply a MC method to compute completeness. 
We mask regions according to the \planck\ 2015 Cosmology mask \citep{Planck_CMB}, along with the point source mask \citep{Planck_pt_src}, and an additional exclusion mask extending 1\degree\ from the center of all clusters in the PSZ2 catalog of P16.
From the remaining unmasked sky regions, we randomly select 100 locations, each separated by at least 5\degree\ in order to sample a representative range of noise environments. At each location, we inject clusters with varying \yclus\ and \thetaclus\ values and determine the associated S/N from the MMF3 algorithm. Our assumed model for the cluster does not include any signal from correlated large-scale structure, and, in general, neither do the random sky locations. Thus, the impact of such structure is not included in the derived S/N. The completeness function is then defined as the detection probability curve, which relates the likelihood of detection above a S/N threshold of 6.5 to \yclus\ and \thetaclus. 

\begin{figure}[!tbp]
  \centering
    \includegraphics[width=0.49\textwidth]{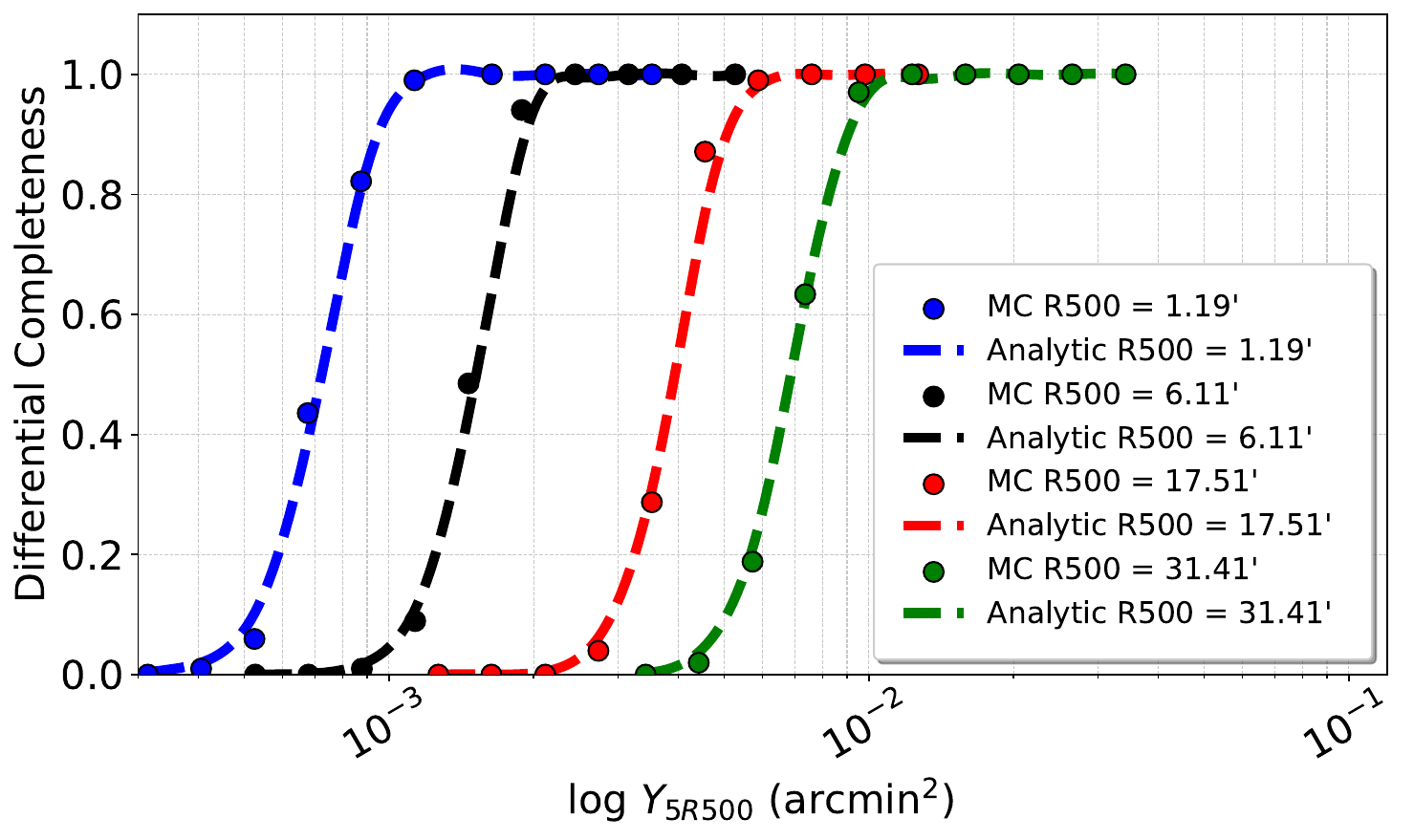}
  \caption{Comparison of our MC completeness averaged over the unmasked sky to the "semi-analytical" ERF completeness for $\theta_{500} = 1.19$\arcmin, 6.11\arcmin, 17.51\arcmin, and 31.41\arcmin, and an S/N threshold of 8.5.}
  \label{fig:completeness_consistency}
\end{figure}

To validate our completeness pipeline, we first perform a direct comparison to the completeness function found by P16 for spherical clusters. We find good consistency, including percent-level agreement with semi-analytical calculations obtained from the error function (ERF)
\begin{equation}
   P(d| Y_{500}, \sigma_Y (\theta_{500}), q) = \frac{1}{2}\left[1 + \text{erf}\left(\frac{Y_{500} - q\sigma_{Y}(\theta_{500})}{\sqrt{2}\sigma_Y (\theta_{500})}\right) \right],
\end{equation}
where $q$ is the detection threshold, and $\sigma_Y(\theta_{500})$ is the estimate of the noise at that angular scale, see Fig.~\ref{fig:completeness_consistency}.

\begin{figure}[!tbp]
  \centering
    \includegraphics[width=0.49\textwidth]{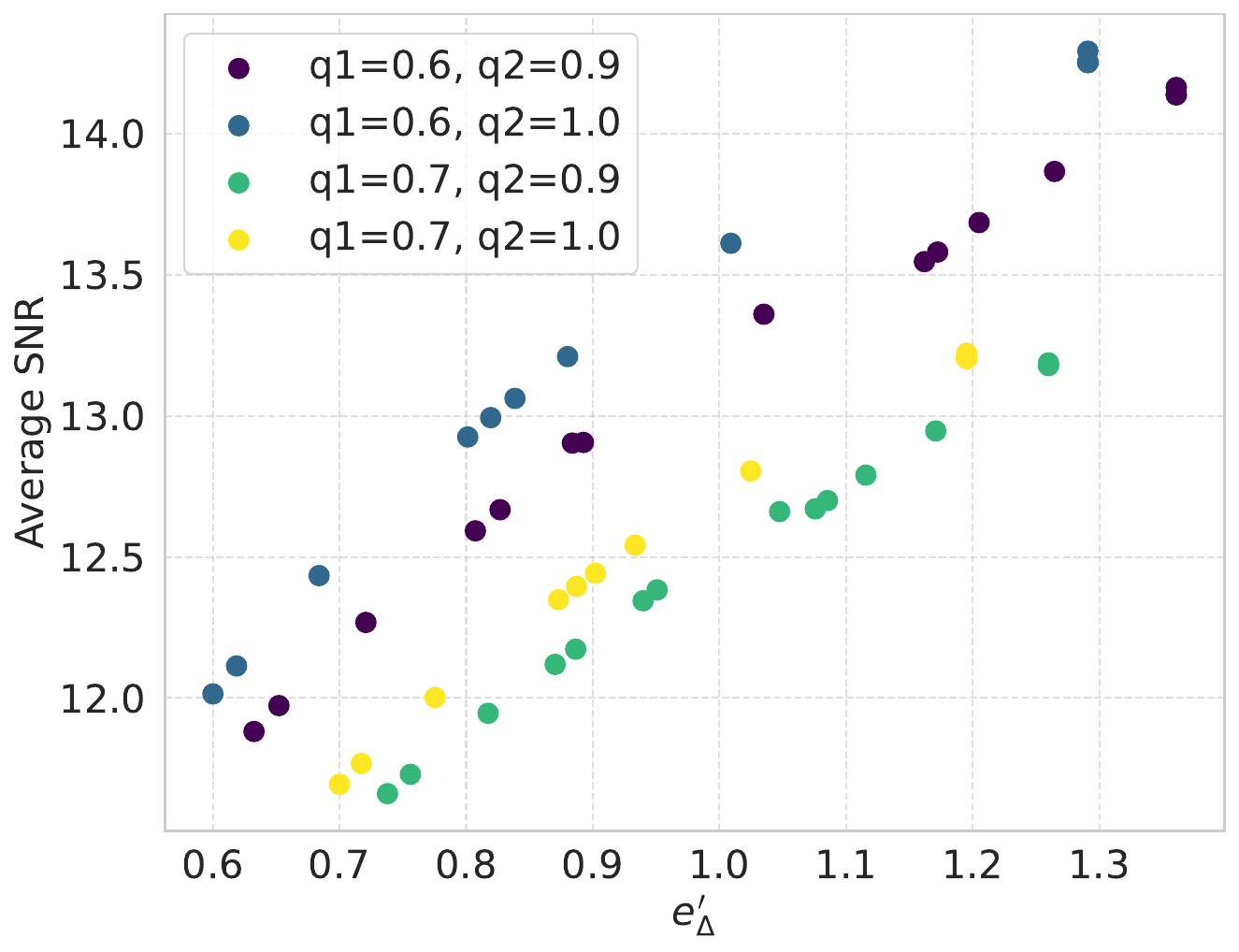}
  \caption{Average MMF3 S/N as a function of $e'_\Delta$ for different triaxial axial ratios, for a clusters with an effective spherical $\theta_{500} = 5.63$\arcmin\ and a fixed value of \yclus.}
  \label{fig:S/N_edelprime}
\end{figure}

Following this validation, we extend our MC completeness calculation to triaxial clusters, constructing completeness curves to quantify how detection probability varies with triaxial axial ratios and orientation. In addition to the masks noted above, these samples include additional constraints: Tier 1 is limited to $\text{decl.} > 0$\degree, while both Tier 1 and Tier 2 have a low \xmm\ visibility mask to exclude regions with visibilities < 55 ks per orbit \citep{CHEXMATE}. From the unmasked sky regions, we again select 100 widely separated map cutouts for both Tier 1 and Tier 2, ensuring good sampling of the available sky. We then insert triaxial clusters with varying axial ratios, orientations, \yclus, and \thetaclus\ into these cutouts and compute completeness by averaging detection probabilities over all sky positions. 
To model triaxial shape and orientation effects on MMF3 S/N, we introduce a new parameter, 
\begin{equation}
    e'_\Delta = \frac{l_{\text{los}}}{({l_{\text{p1}}*l_{\text{p2}}})^{0.5}},
\end{equation}
where, as before, $l_{los}$ is the semi-major axis of the ellipse along the line of sight, and $({l_{\text{p1}}*l_{\text{p2}}})^{0.5}$ is the effective spherical radius of the projected ellipse on the sky, see Fig. \ref{fig:triax_ellip}. We find that $e'_{\Delta}$ exhibits a complex dependence on $q_1$,\ $q_2$,\ $\cos\theta$, and $\phi$, but correlates monotonically with S/N—higher values of $e'_{\Delta}$ correspond to higher S/N for a fixed $(q_1, q_2)$, as shown in Fig. \ref{fig:S/N_edelprime}. We have established this correlation for the full range of geometries considered in this paper, but we only show a subset in Fig. \ref{fig:S/N_edelprime} for clarity. Thus, rather than using two separate orientation parameters $(\cos\theta, \phi)$, we consolidate them into $e'_{\Delta}$ for each $q_1$,\ $q_2$. Using this framework, we construct a comprehensive lookup table that quantifies detection probability as a function of SZ parameters and triaxial shape, thus describing $P(d|Y_{500}, \theta_{500}, q_1, q_2, e'_{\Delta})$. 

\begin{figure*}[!tbp]
  \centering
    \includegraphics[width=\textwidth]{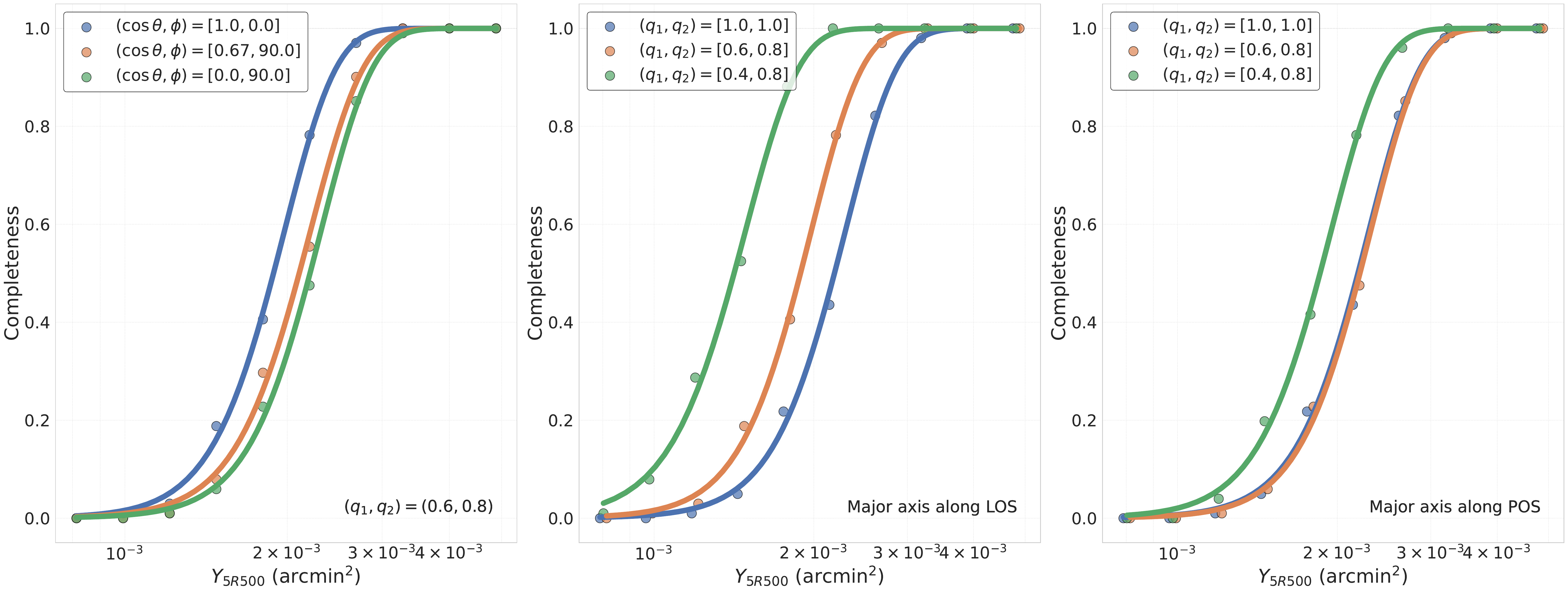}
  \caption{ Completeness as a function of \yclus\ for a cluster with an effective spherical $\theta_{500} = 5.63$\arcmin. \textbf{Left:} Completeness for different orientations for a cluster with $q_1 = 0.6, q_2 = 0.8$. \textbf{Middle: } Completeness for a cluster with different triaxialities but with major axis oriented along LOS with $\cos\theta = 1.0, \ \phi = 0^\circ$. \textbf{Right: } Completeness for a cluster with different triaxialities but with major axis oriented on the plane of sky (POS) with $\cos\theta = 0.0, \ \phi = 90^\circ$.}
  \label{fig:triax_comp_slice}
\end{figure*}
We present three slices of the five-parameter completeness function in Fig. \ref{fig:triax_comp_slice}, illustrating completeness as a function of triaxial axial ratios and orientations. We find a clear dependence on orientation, where clusters elongated along the LOS exhibit a higher probability of detection for the same \yclus, \thetaclus, and shape parameters. Additionally, we observe that for the same effective spherical \thetaclus\ but for other orientations, there is no clear trend between detectability and cluster triaxiality. We thus find that the primary factor influencing detectability is the extent of the cluster along the LOS. For triaxial clusters where the major axis is not closely aligned with the LOS, varying triaxialities can still produce more, or less, centrally peaked SZ profiles relative to spherical clusters due to projection. Consistent with the results reported by \citet{Gallo2024}, we find that the more centrally peaked clusters from this projection effect are more easily detected.

To ensure a well-sampled and smoothly varying lookup table to connect detection probability with the relevant parameters, we fit the completeness function for each ($\theta_{500}, q_1, q_2, e'_\Delta$) to an error function (ERF) of the form:
\begin{equation}
    P(d) = \frac{1}{2}\left[1 + \text{erf}\left(\frac{Y_{500} - q*a(R_{500}, q_1, q_2, e'_\Delta)}{\sqrt{2}*b(R_{500}, q_1, q_2, e'_\Delta)}\right) \right].
    \label{eqn:ERF_tri}
\end{equation}
Relative to the ERF in P16, this form incorporates modified mean ($a$) and variance ($b$) terms to capture non-Gaussian deviations introduced by cluster triaxiality. We then perform a grid interpolation over our lookup table to determine $a$ and $b$ as a function of $q_1$, $q_2$, and $e'_\Delta$. 

\section{Mock CHEXMATE Catalogs} \label{mock_CHEX}
\begin{figure*}[h]
  \centering
    \includegraphics[width=\textwidth]{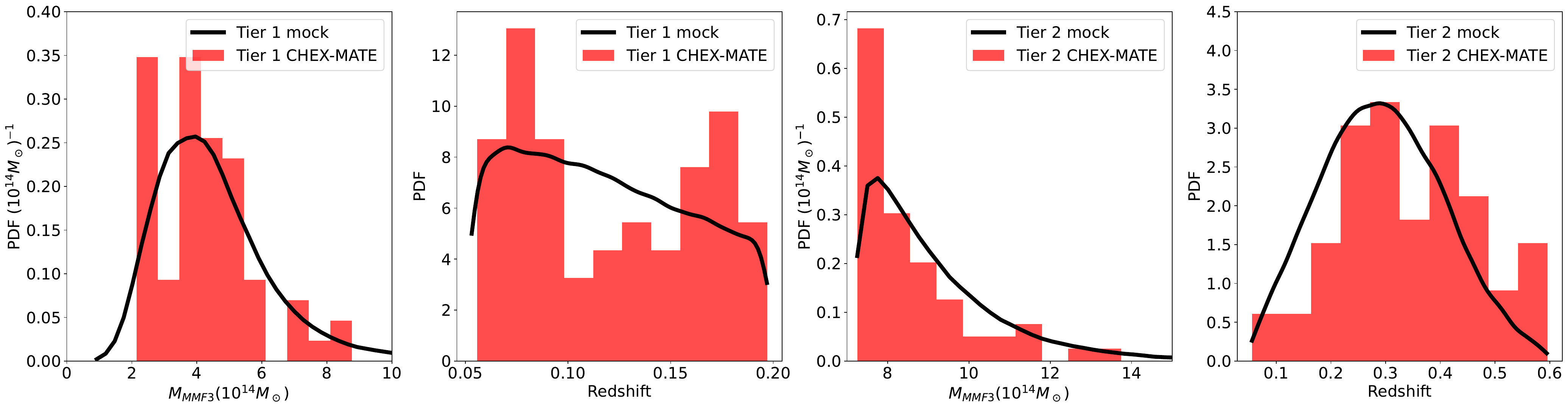}
  \caption{Mass and redshift distributions for the observed Tier 1 and Tier 2 \chexmate\ samples compared to the kernel density estimation (KDE) distribution of those obtained from our 1000 mock \chexmate\ selections. The distributions are consistent, indicating our mock samples accurately reproduce the masses and redshifts of the observed sample.} 
  \label{fig:Tier1_2_Mz_dist}
\end{figure*}

We use the triaxial completeness given by the ERF in Eqn.~\ref{eqn:ERF_tri} to construct mock \chexmate\ catalogs, allowing us to quantify the on-average distribution of orientations and geometries that arise from this sample selection. First, we generate mock universes with halo populations sampled from the \citet{Tinker_2008} HMF assuming the \citet{Planck2018} cosmology, considering redshifts $0 \leq z \leq 0.6$ and $M_{500}$ masses between $10^{14} M_\odot$ and $10^{16} M_\odot$, while also applying the sky fraction constraints for both Tier 1 and Tier 2 samples.
We convert the generated halo masses ($M_{500}$) and redshifts ($z$) into $\theta_{500}$ and $Y_{500}$ using the established scaling relations from \citet{PSZ1Cosmo}:
\begin{align}
&E^{-\beta}(z)\left[\frac{D_{\mathrm{A}}^2(z) \bar{Y}_{500}}{10^{-4} \mathrm{Mpc}^2}\right]=Y_*\left[\frac{h}{0.7}\right]^{-2+\alpha}\left[\frac{(1-b) M_{500}}{6 \times 10^{14} \mathrm{M}_{\odot}}\right]^\alpha
\label{eqn:SR1}
\end{align}
and
\begin{align}
&\bar{\theta}_{500}=\theta_*\left[\frac{h}{0.7}\right]^{-2 / 3}\left[\frac{(1-b) M_{500}}{3 \times 10^{14} \mathrm{M}_{\odot}}\right]^{1 / 3} E^{-2 / 3}(z)\left[\frac{D_{\mathrm{A}}(z)}{500 \mathrm{Mpc}}\right]^{-1}
\label{eqn:SR2}
\end{align}
where $E(z) = H(z)/H_0$, $D_A (z)$ is the angular diameter distance, and $(1-b)$ is the mass bias factor, which is equal to 0.62 when requiring consistency with the cosmology obtained by \citet{Planck2018}. The values of the scaling and normalization parameters are obtained from \cite{PSZ2Cosmo}. We then perform volume selections to form mock Tier $1$ ($0.05<z<0.2$) and Tier $2$ ($z<0.6$) catalogs from halos generated using the HMF. To assign triaxial shapes and orientations to halos, we randomly draw axial ratio values for the gas from the empirical distributions obtained from the IllustrisTNG simulations presented in
\citet{triax_prior}, along with random draws for $\cos\theta \in [$0, 1] and $\phi \in [$0\degree, 90\degree]. 
Then, using the calibrated ERF, we compute the completeness for each halo and include it in the resulting catalog based on a probabilistic sampling. Specifically, for each halo a uniform random value between 0 and 1 is drawn, and the halo is included in the catalog if the value of $P(d)$ is larger than this random value. We repeat this process for $1000$ mock universes for both Tier 1 and Tier 2.
\\\\
Since Tier 2 requires an additional selection based on MMF3-derived mass, we reprocess those completeness-selected clusters through the MMF3 algorithm. This provides the extracted $Y_{500}$ for each cluster at every filter scale \thetaclus\ of MMF3. From this $Y-\theta$ relation we then determine the value of \mmmf\ that simultaneously satisfies Eqns.~\ref{eqn:SR1} and \ref{eqn:SR2}.
Since this process is repeated across 100 distinct map cutouts, we obtain 100 different values of \mmmf\ for each cluster, reflecting the variation in recovered mass due to noise fluctuations. To assign a single value of \mmmf\ to each cluster, we perform a probabilistic selection based on the cluster’s detection completeness. Specifically, we randomly select a value of \mmmf\ from among the $N$ highest values obtained from the 100 cutouts, where $N = P(d) \times 100$.
This method effectively models the S/N-based mass boosting, as clusters with lower intrinsic S/N values are only included in the catalog when they fall within favorable map cutouts that result in boosted S/N values. Following \chexmate, the final Tier 2 selection is then performed by requiring $\text{M}_{\text{MMF3}} > 7.25 \times 10^{14} M_\odot$. 
\\\\
As an additional validation of our detection pipeline, we find that the average number of halos detected across 1,000 mock realizations for both the Tier 1 and Tier 2 selections is 59, closely matching the 61 clusters observed in the actual \chexmate\ sample. Furthermore, as shown in Fig. \ref{fig:Tier1_2_Mz_dist}, the mass and redshift distributions of the mock catalogs are consistent with the observed distributions when accounting for Poisson fluctuations due to the relatively small sample size.
For Tier 1, the $p$-values from a Kolmogorov-Smirnov (KS) test of the mass and redshift distributions are $0.43$ and $0.29$, indicating very good agreement. For Tier 2, the $p$-values are lower, equal to $0.011$ for both the mass and redshift distributions, due primarily to the relative excess of high mass and low redshift systems detected in the mocks. While these $p$-values suggest possible discrepancies at low statistical significance, we note that \citet{PSZ2Cosmo} also found a similar excess of low-$z$ clusters predicted by the best-fit HMF relative to their abundance measurement from the \planck\ cluster catalog. We thus conclude that our Tier 2 mocks provide a reasonable description of the observed \chexmate\ sample.

\begin{figure*}[!tbp]
  \centering
    \includegraphics[width=0.85\textwidth]{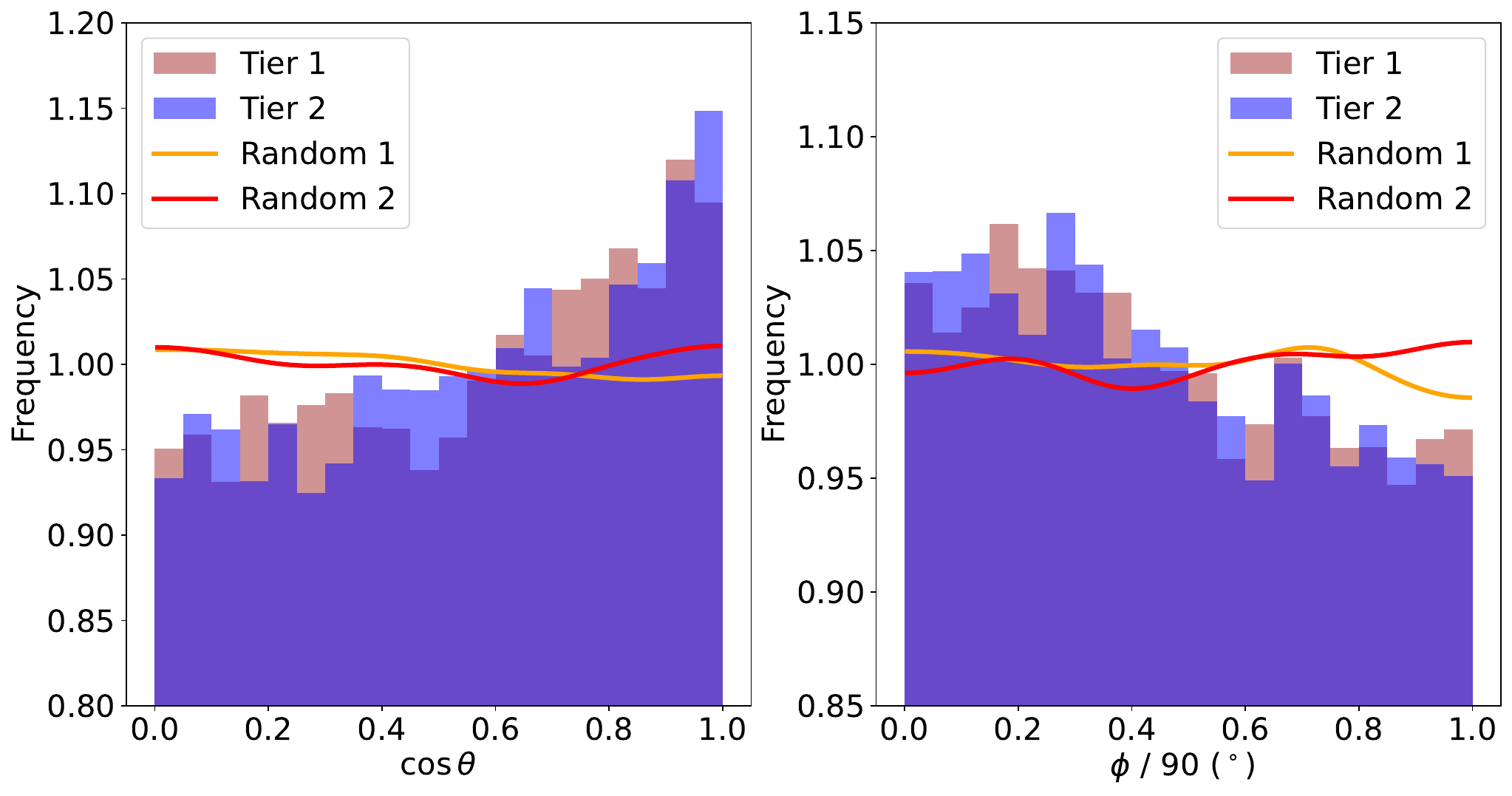}
  \caption{Distributions of orientation angles $\cos\theta$ and $\phi$ for the mock Tier 1 and Tier 2 samples, along with randomly selected samples with similar distributions in mass and redshift. 
  There is a clear bias toward larger values of $\cos\theta$ and smaller values of $\phi$ for the \chexmate\ samples compared to the random samples, indicating that an SZ selection results in a preference for clusters elongated along the LOS. }
  \label{fig:Tier1_2_cos_phi_dist}
\end{figure*}

To assess the triaxial geometry of a \chexmate-selected sample, Fig. \ref{fig:Tier1_2_cos_phi_dist} shows the distributions of $\cos\theta$ and $\phi$ obtained for the Tier 1 and Tier 2 samples. For comparison, the values of  $\cos\theta$ and $\phi$ obtained from randomly selected halos matching the mass and redshift distributions of the \chexmate-selected samples are also shown, and, as expected, are consistent with random orientations.
Within the \chexmate-selected samples, we see a clear excess in clusters that are elongated along the LOS, with a preference for halos with $\cos\theta$ values close to 1 and $\phi$ values close to 0\degree,
implying that an SZ selection results in a population biased toward LOS elongation. Quantitatively, the $p$-value obtained from a KS test comparing the distributions of $e'_\Delta$ for the \chexmate-selected and random samples is $2.7 \times 10^{-30}$, suggesting they are inconsistent at high statistical significance. We note that, while the excess of clusters with $\cos\theta$ values near 1 is easy to understand, since it implies that the major axis is well aligned with the LOS, the excess of clusters with $\phi$ values close to 0\degree\ is more subtle. It arises because, when the major axis is well aligned with the POS (i.e., $\cos\theta$ near 0), values of $\phi$ close to 0\degree\ imply that the intermediate axis is well aligned with the LOS. Such clusters thus have a boosted SZ signal compared to those with $\phi$ close to 90\degree, which have their minor axis well aligned with the LOS.

\section{Weak Lensing Mass Bias} \label{WL_mass_bias}
\begin{figure*}[!tbp]
  \centering
  \begin{subfigure}{\textwidth}
    \centering
    \includegraphics[width=\textwidth]{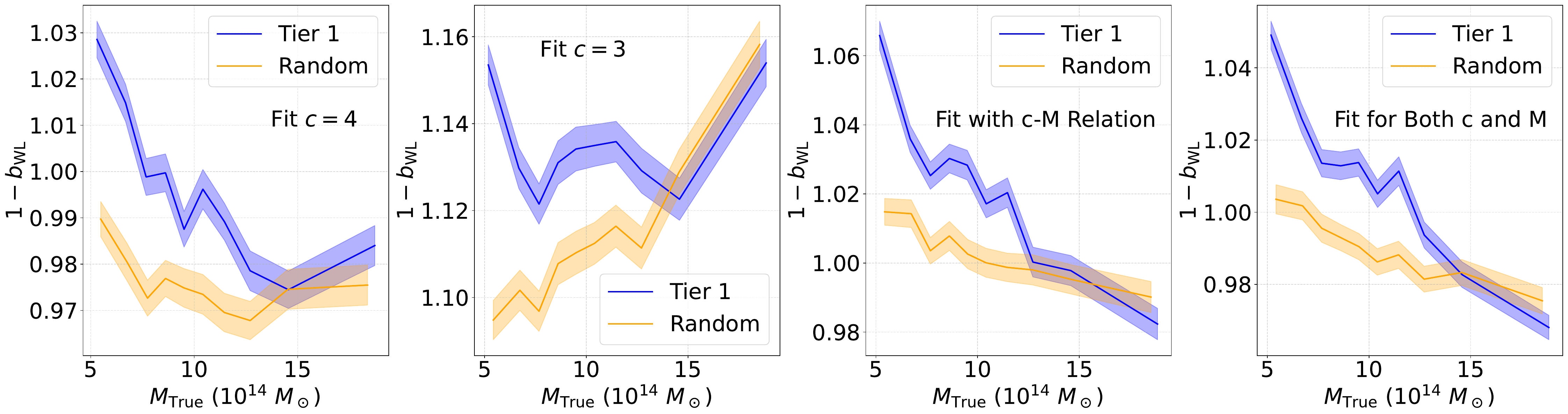}
  \end{subfigure}
  \vspace{1em} 
  \begin{subfigure}{\textwidth}
    \centering
    \includegraphics[width=\textwidth]{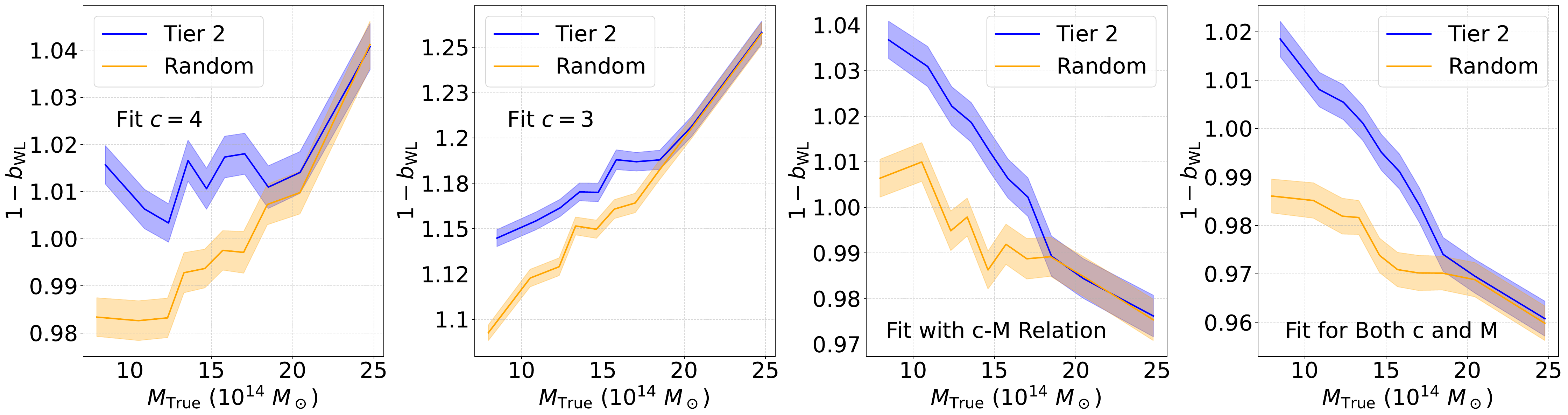}
  \end{subfigure}
  \caption{Comparison of \wlbias\ as a function of true mass for the mock \chexmate\ Tier 1 (\textbf{Top}) and Tier 2 (\textbf{Bottom}) selected samples (purple lines and shaded 68\% confidence level regions), compared to a random distribution of halos with matched mass and redshift distributions (orange lines and shaded 68\% confidence level regions) with varying priors on the value of $c_{200}$ used to obtain \mwl\ from the mock shear data. \textbf{Left: } $c_{200} = 4$. \textbf{Middle Left: }  $c_{200} = 3$. \textbf{Middle Right: } $c_{200}$ obtained from the $c$--$M$ relation of \citet{Diemer_2019}. \textbf{Right: } $c_{200}$ varied as a free parameter.}
  \label{fig:Combined_WL_Bias}
\end{figure*}

To assess whether the bias in orientation from an SZ selection results in a measurable \mwl\ bias, we compare \mwl\ estimates obtained from clusters detected in the mock CHEX-MATE catalogs to randomly selected samples of clusters with similar distributions in mass and redshift. We generate mock shear maps for each mock cluster from its shape parameters, mass, and redshift, following Gavidia et al. (in prep). In brief, we assume that the gravitational potential follows the geometry of the ICM, which is described by a triaxial ellipsoid. Specifically, we adopt the same axial ratios and orientation for the gravitational potential as those used for the gas distribution when generating the mock SZ images. This assumption is well motivated by simulations \citep[e.g.,][]{Lau2011}. An NFW density profile \citep{Navarro1997} in a triaxial basis is used to model the matter density distribution within the cluster, and a numerical inversion of Poisson's equation is performed to calculate the gravitational potential. The 3D gravitational potential is then projected to obtain the lensing potential, and partial derivatives of the model lensing potential are taken to obtain the model shear maps. As with the values of \yclus\ and \thetaclus, the concentration $c_{200}$ and total mass M$_{200}$ (defined as M$_{\text{True}}$ hereafter) for these triaxial distributions are computed by spherically-averaging the 3D halo. For each halo, the values of the spherically-averaged $c_{200}$ and M$_{\text{True}}$ are randomly drawn from the relation given by \cite{Diemer_2019}, assuming that $\log(c_{200})$ has an intrinsic scatter of 0.16. To isolate the impact of projection effects on these idealized mass estimates, we introduce a minimal level of uniform noise to the shear maps, approximately four orders of magnitude below current observational sensitivities.

These mock shear maps are then fitted to a spherically-symmetric NFW model, analogous to the standard approach utilized for cosmological analyses, performed via a least-squares approach. Four different fits are performed using different values of $c_{200}$, including fixing its value to either 3 or 4, fixing its value based on the concentration-mass relation from \citet{Diemer_2019}, and allowing its value to float (within the range between 1 to 10). We perform these fits for the mock Tier 1 and Tier 2 samples, along with the randomly selected samples matching their redshift and mass distributions. We then compute the average ratio of the fitted \mwl\ and the true halo mass, which we define as $1-b_{\text{WL}} = \text{M}_{\text{WL}} / \text{M}_{\text{True}}$, within discrete mass bins for all cases, see Fig.~\ref{fig:Combined_WL_Bias}. For both Tier 1 and Tier 2, the \mwl\ of the \chexmate-selected samples are higher than those obtained from the random samples at the low-mass end, trending toward agreement at the high-mass end. This is as expected, since the up-scatter in S/N due to clusters elongated along LOS is more pronounced in lower mass (and thus lower S/N) clusters, with higher mass clusters generally having a higher S/N which allows for detection irrespective of orientation. Further, for $c=3$ and $c=4$, we see a sharp increase in \mwl\ for both the random and \chexmate-selected samples at the high mass end. This is a result of utilizing the relation of \citet{Diemer_2019} when generating the halos, as they found an increase in $c_{200}$ at the high mass end.
\\

\begin{figure*}[!tbp]
  \centering
  \begin{subfigure}[b]{0.49\textwidth}
    \includegraphics[width=\textwidth]{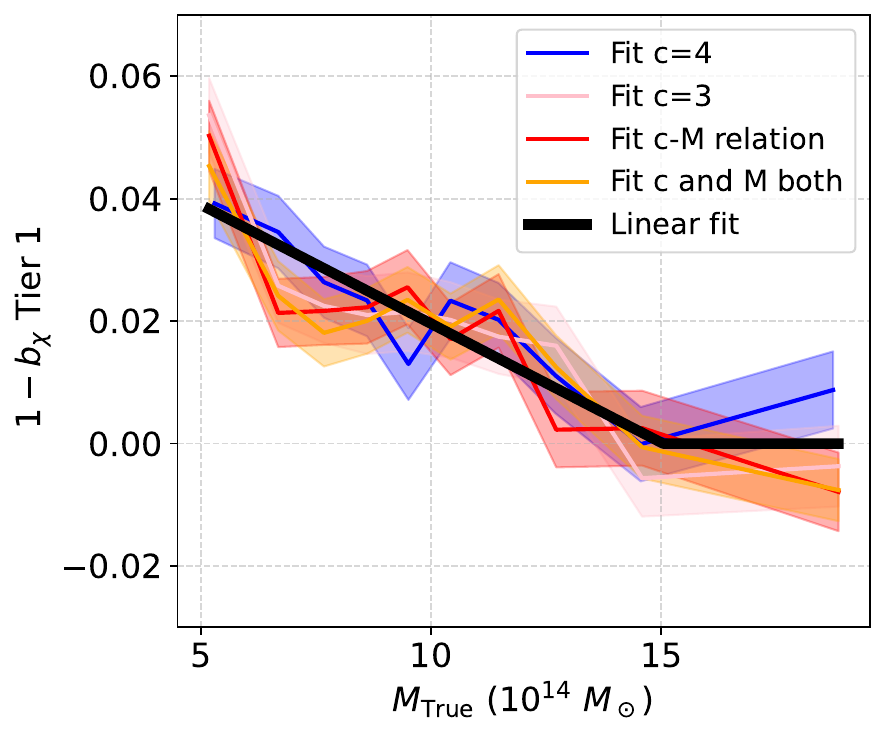}
  \end{subfigure}
  \hfill
  \begin{subfigure}[b]{0.49\textwidth}
    \includegraphics[width=\textwidth]{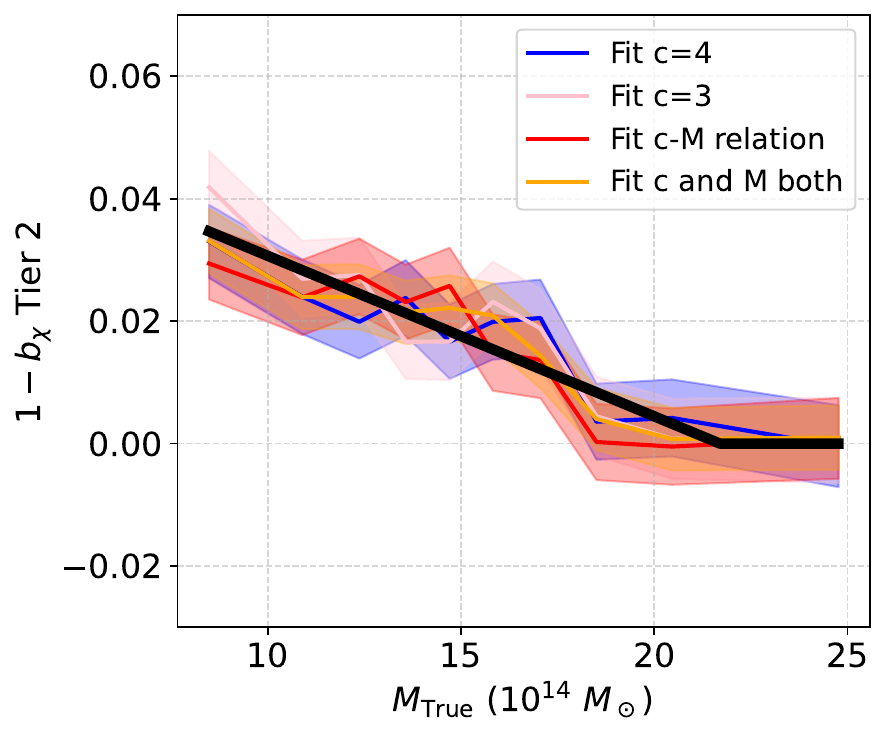}
  \end{subfigure}
  \caption{The ratio between \wlbias\ obtained from the \chexmate-selected samples (\textbf{Left:} Tier 1, \textbf{Right:} Tier 2), and \wlbias\ obtained from random samples with matching mass and redshift distributions, defined as \obias\ . This plot thus isolates the bias due solely to orientation from an SZ-selected sample. This orientation bias is approximately independent of fitting methodology, as illustrated by the good agreement between all four choices utilized in this work. Overlaid in black is a broken linear fit, which reproduces the derived values of \obias\ to within an rms scatter of 0.003.}
  \label{fig:Tier_WL_bias_combine}
\end{figure*}

We note that the overall value of \wlbias\ depends strongly on our fitting methodology, i.e., the assumed prior on $c_{200}$, as expected due to the strong anti-correlation between c and M. However, the ratio of \wlbias\ obtained from the \chexmate-selected and random samples, which we define as \obias\, appears to be independent of fitting methodology, as illustrated in Fig.~\ref{fig:Tier_WL_bias_combine}. In other words, the portion of \wlbias\ associated with the on-average orientation of an SZ selection is robust to fitting choice. For both Tier 1 and Tier 2, we find that \obias\ is approximately 4\% at the low-mass end, with an overall average value near 2\%.
Thus, for the \chexmate\ samples, and likely any \planck\ SZ-selected sample, the correction for this bias in \wlbias\ due to orientation should be universal regardless of WL fitting methodology. To facilitate such a correction, we provide the parameters of a linear regression fit of the form
\begin{align*}
1-b_\chi  =
&\begin{cases}
    k \times (\text{M}_{\text{True}} - \text{M}_{\text{break}}) \ / \ 10^{14} \ \text{M}_\odot & \text{if }  \text{M}_{\text{True}} < \text{M}_{\text{break}} \\
    0 & \text{if } \text{M}_{\text{True}} \geq \text{M}_{\text{break}}
\end{cases}
\end{align*}
 to describe selection-based orientation-induced bias in \mwl\ for both the Tier 1 and Tier 2 samples, with the results summarized in Tab.~\ref{tab:Tier12_combined_bias}. We find that these linear fits accurately reproduce the derived bias to within 0.3\%, sufficient for per cent-level mass calibration. Because of the different mass and redshift ranges for Tier 1 and Tier 2, along with the additional selection based on MMF3-derived mass for Tier 2, there is no expectation that the fitted slope and break point should be consistent for the two samples, and we indeed find modest differences in their values.
\begin{table}[ht]
\centering
\begin{tabularx}{0.49\textwidth} { 
  | >{\raggedright\arraybackslash}X 
| >{\raggedright\arraybackslash}X 
  | >{\raggedright\arraybackslash}X | }
 \hline
 \chexmate\ & M$_{\text{break}}$ & $k$  \\
 \hline
 Tier 1 & $15.03 \times 10^{14}$ M$_{\odot}$ & $-0.0039 $ \\ 
 \hline
 Tier 2 & $21.67 \times 10^{14}$ M$_{\odot}$ & $-0.0026$ \\ 
 \hline
\end{tabularx}
\caption{Parameters of the linear fit of \obias\ for the Tier 1 and Tier 2 \chexmate\ samples, where \obias\ gives the bias in \mwl\ due solely to the orientation bias from the \planck\ SZ selection.}
\label{tab:Tier12_combined_bias}
\end{table}

\section{Conclusions} \label{conc}
In this paper, we investigate the impact of triaxiality and orientation on SZ selection and the associated \mwl\ values for such a sample, focusing on the specific example of \chexmate. Quantifying these selection effects is crucial for future cluster cosmology studies, as it enables corrections to the \mwl\ values utilized for calibration of the halo masses. 
\\\\
To obtain this goal, we recreated the MMF3 algorithm used by P16 for the detection of clusters based on their SZ signal in the \planck\ all-sky maps. Using a MC method based on the injection of mock clusters in randomly sampled cutouts from \planck\ HFI maps within the unmasked region of the sky, we confirmed that the values of S/N and \yclus\ obtained from our implementation of the MMF3 are consistent with those obtained by P16 for spherically-symmetric clusters.
By then applying this algorithm to mock observations of clusters with triaxial shapes, we determined the completeness as a function of SZ signal strength, angular size, and triaxial shape and orientation.  
\\\\
Using our measured triaxial completeness, we construct mock \chexmate\ catalogs to quantify selection biases introduced by triaxiality and orientation for SZ-detected clusters. We find a clear orientation bias, where clusters elongated along the LOS are preferentially detected due to their centrally boosted SZ signals. By fitting \mwl\ to the mock Tier 1 and Tier 2 detected clusters, along with randomly selected clusters with comparable mass and redshift distributions, we identify a mass-dependent \mwl\ bias due solely to the shape and orientation of the SZ selected samples, \obias. This bias is as large as $4\%$ in the lowest mass clusters, and approximately $2\%$ when averaged across the full mass range. In addition, \obias\ appears to be independent of the exact WL fitting methodology. We also note that the value of \obias\ we find is expected to be a lower limit, since we do not account for correlated LSS in our selection or WL modeling. Such structure is expected to occur preferentially along the elongation direction of the halo, and it would thus boost the SZ signal when viewed in that orientation (resulting in a stronger preference for LOS elongation in the \chexmate-selected samples), along with boosting the value of \mwl\ at fixed LOS elongation. 

We also consider the implications of our results in the context of cluster cosmology from halo abundance measurements. First, we have found that \mwl\ is systematically overestimated in SZ selected samples from \planck, and so the true masses are on-average lower than WL-derived masses for such a sample. The discrepancy between halo abundance and primary CMB measurements from \planck\ requires higher halo masses for the cluster sample, and so this correction goes in the opposite direction. In addition, we note that the on-average $2\%$ systematic overestimate of \mwl\ that we measure is subdominant to current systematic uncertainties on \mwl\ calibration,
and thus it does not significantly impact previously published results. 
However, upcoming surveys like CMB-S4 and \euclid\ require percent-level mass calibration \citep{Raghunathan2022,euclid}, and so such biases must be accounted for in their analyses. 

\begin{acknowledgements}
This research was supported by the International Space Science Institute (ISSI) in Bern, through ISSI International Team project \#565 ({\it Multi-Wavelength Studies of the Culmination of Structure Formation in the Universe}). H.S, J.S, A.G and J.K were supported by NASA Astrophysics Data Analysis Program (ADAP) Grant 80NSSC21K1571. SE, MS, MR acknowledges the financial contribution from the contracts Prin-MUR 2022 supported by Next Generation EU (M4.C2.1.1, n.20227RNLY3 {\it The concordance cosmological model: stress-tests with galaxy clusters}). MS acknowledges the financial contributions from contract INAF mainstream project 1.05.01.86.10 and INAF Theory Grant 2023: Gravitational lensing detection of matter distribution at galaxy cluster boundaries and beyond (1.05.23.06.17). SE, MR acknowledges the financial contributions from the European Union’s Horizon 2020 Programme under the AHEAD2020 project (grant agreement n. 871158). GWP acknowledges long-term support from CNES, the French space agency. LL acknowledges the financial contribution from the INAF grant 1.05.12.04.01. EP acknowledges support from CNES, the French national space agency and from ANR the French Agence Nationale de la Recherche, under grant ANR-22-CE31-0010. J.K acknowledges the support by Basic Science Research Program through the National Research Foundation of Korea (NRF) funded by the Ministry of Education(2019R1A6A1A10073887) and the 2025 KAIST-U.S. Joint Research Collaboration Open Track Project for Early-Career Researchers, supported by the International Office at the Korea Advanced Institute of Science and Technology (KAIST). M.G. acknowledges support from the ERC Consolidator Grant \textit{BlackHoleWeather} (101086804). MDP acknowledges financial support from PRIN-MUR grant 20228B938N {\it"Mass and selection biases of galaxy clusters: a multi-probe approach"} funded by the European Union Next generation EU, Mission 4 Component 1 CUP C53D2300092 0006. DE acknowledges support from the Swiss National Science Foundation (SNSF) through grant agreement \#200021\_212576.

\end{acknowledgements}

%
%

\bibliographystyle{aa} 
\bibliography{bibliotech} 

\end{document}